\documentclass{ws-ijmpcs}

\begin{document}

\newcommand{\lec}{\mathrm{e}}
\newcommand{\ionperso}{\mathrm{i}}
\newcommand{\therm}{\mathrm{th}}
\newcommand{\dif}{\mathrm{d}}

\markboth{M.~Melzani, R.~Walder, D.~Folini, C.~Winisdoerffer}
{Differences Between Real and Particle-In-Cell Plasmas}

%
\catchline{}{}{}{}{}
%

\title{DIFFERENCES BETWEEN REAL AND PARTICLE-IN-CELL PLASMAS: EFFECTS OF COARSE-GRAINING}

\author{MICKA\"EL MELZANI}
\address{mickael.melzani@ens-lyon.fr}

\author{ROLF WALDER}
\address{rolf.walder@ens-lyon.fr}

\author{DORIS FOLINI}
\address{doris.folini@ens-lyon.fr}

\author{CHRISTOPHE WINISDOERFFER}
\address{cwinisdo@ens-lyon.fr}

\address{\'Ecole Normale Sup\'erieure de Lyon, Centre de Recherche Astrophysique de Lyon, UMR CNRS 5574, Universit\'e de Lyon\\
Lyon, France\\
}

\maketitle

\begin{history}
\received{Day Month Year}
\revised{Day Month Year}
\end{history}

\begin{abstract}
The PIC model relies on two building blocks. The first stems from the capability of computers to
handle only up to $\sim10^{10}$ particles, while real plasmas contain from $10^4$ to $10^{20}$ particles per Debye
sphere: a coarse-graining step must be used, whereby of the order of $p\sim10^{10}$ real particles are
represented by a single computer superparticle. The second is field storage on a grid with its subsequent
finite superparticle size. We introduce the notion of coarse-graining dependent quantities, i.e. physical
quantities depending on the number $p$. They all derive from the plasma parameter $\Lambda$, which we show to
be proportional to $1/p$. 

We explore three examples: the rapid collision- and fluctuation-induced thermalization of plasmas with different temperatures, 
that scale with the number of superparticles per grid cell and are a factor $p\sim10^{10}$ faster than in real plasmas;
the high level of electrostatic fluctuations in a thermal plasma, with corrections due to the finite superparticle sizes;
and the blurring of the linear spectrum of the filamentation instability, where the 
fastest growing modes do not dominate the total energy because of a high level of fluctuations. 

We stress that the enhanced collisions and correlations of PIC plasmas must be kept negligible toward kinetic physics.
\end{abstract}

\ccode{PACS numbers: 11.25.Hf, 123.1K}

\section{Introduction}

Particle-in-cell simulations have brought tremendous new
insights into collisionless astrophysical plasmas, for example regarding instabilities, kinetic turbulence, 
or particle acceleration in shocks and in magnetic reconnection.
However, there remain a number of questions concerning the degree to which PIC
models are able to completely mirror real plasmas. 
In this manuscript, and in more detail in Ref.~\refcite{mmelzani}, we address the incidence of coarse-graining, i.e., of the 
fact that each computer particle in the PIC plasma actually represents many real plasma particles.
The need for coarse-graining becomes evident when looking at the number of plasma particles per Debye sphere,
given by the plasma parameter $\Lambda$. In a real plasma $\Lambda$ ranges from $10^4$ to $10^{20}$
(e.g., $\Lambda\sim10^{6}$ in solar coronal loops; $10^{12}$ in the magnetotail, 
magnetopause, or in typical Crab flares; $10^{17}$ in AGN jets),
while in computer experiments, where we have to simulate thousands
to millions of Debye spheres, $\Lambda$ reaches hardly a few tens.
The corresponding number of particles per computer particles -- that we will denote as \textit{superparticles} --
then reaches $p\sim 10^3$ to $10^{19}$.

\section{Principles of the PIC Algorithm}

A PIC code solves Maxwell equations with current and charge densities computed from 
the superparticles, and the equations of motion with the Lorentz force for each superparticle.
The fields are stored on a grid, while the superparticles evolve continuously in position and velocity space.

Information between the superparticles inside the cells and the fields stored at grid nodes is communicated via interpolation,
and this interpolation is equivalent to considering the superparticles as clouds of charge with a finite extent\cite{Birsdall1985,Hockney1988}. 
This in turn implies a vanishing two-point force between superparticles at short distances, and thus reduces drastically 
the influence of collisions. It helps the PIC plasma to remain collisionless even if its plasma parameter is quite low\cite{Birsdall1985}.
Discretization also brings numerical stability issues, that have been explored in depth in Refs.~\refcite{Birsdall1985,Hockney1988}.

\section{Fluid versus Coarse-Graining Dependent Quantities}

We assume that we model a real plasma composed of $N$ real particles by using in the code $N/p$ superparticles.
The number of real particles represented by each superparticle is thus $p$. 
We also denote by $\rho_\mathrm{sp}^0$ the initial number of superparticles per grid cell in a simulation,
and by $X_0$ the physical size of a cell.

The set of equations solved by the PIC code is not invariant under coarse-graining 
because there are physical quantities that depend on $p$. Such quantities will be termed 
\textit{coarse-graining dependent quantities}. 
By contrast, physical parameters that do not depend on $p$ will be refereed to as \textit{fluid quantities}.
All the fluid physics is thus accurately described by the PIC plasma, 
while some care is to be taken for the coarse-graining dependent physics.

Fluid quantities are invariant under the change of variables $(m,q,n) \rightarrow (m\times p, q\times p, n/p)$,
which amounts to pass from the real plasma of particles of mass $m$, charge $q$, number density $n$, to the plasma of 
superparticles. Examples include the plasma pulsation $\omega_\mathrm{p}^2 = nq^2/(\epsilon_0 m)$,
the thermal velocity $v_\mathrm{th}$, the Debye length $\lambda_{D} = v_\mathrm{th}/\omega_\mathrm{p}$,
and more generally any quantity derived from fluid theory (fluid theories include MHD, two-fluid models, Vlasov-Maxwell system).

The prototype of coarse-graining dependent quantities is the plasma parameter $\Lambda = 4\pi n_\mathrm{e} \lambda_{D\mathrm{e}}^3$.
Since $\Lambda$ is proportional to the number of electrons per Debye sphere, and since the Debye length is $p$-independent, 
the plasma parameter of the PIC plasma reads
\begin{equation}\label{equ:Lambda_PIC}
 \Lambda^\mathrm{PIC}_p = \Lambda/p,
\end{equation}
with $\Lambda = \Lambda^\mathrm{PIC}_{p=1}$ the real plasma parameter.
Another way of seeing this is to recall that the 
plasma parameter $\Lambda$ is the ratio of the particles' kinetic energy to their electrostatic potential energy of interaction and,
as such, varies as $1/p$ because kinetic energy is proportional to the superparticles' mass $m_\mathrm{sp} \propto p$ 
while charge interaction energy involves their charge $q_\mathrm{sp}^2 \propto p^2$. 
Other coarse-graining dependent quantities can be built from $\Lambda$: the thermalization time or the slowing down time 
of fast particles, both due to collisions and correlations, scale as $t_\mathrm{th} \propto \Lambda/\omega_\mathrm{p}$, 
or the level of electrostatic fluctuations $\varepsilon \propto 1/\Lambda$.
Other coarse-graining dependent parameters can be found. Any number of particles per fluid volume will have the same dependency as Eq.~\ref{equ:Lambda_PIC},
for example $n_e(c/\omega_\mathrm{p})^3$.

In the following, we explore situations where coarse-graining effects have visible and important consequences.

\section{Coarse-Graining and Thermalization Times}

\label{Sec_plasma_behavior_thermalization_time}

\begin{figure}
 \centering
 \includegraphics[width=\textwidth]{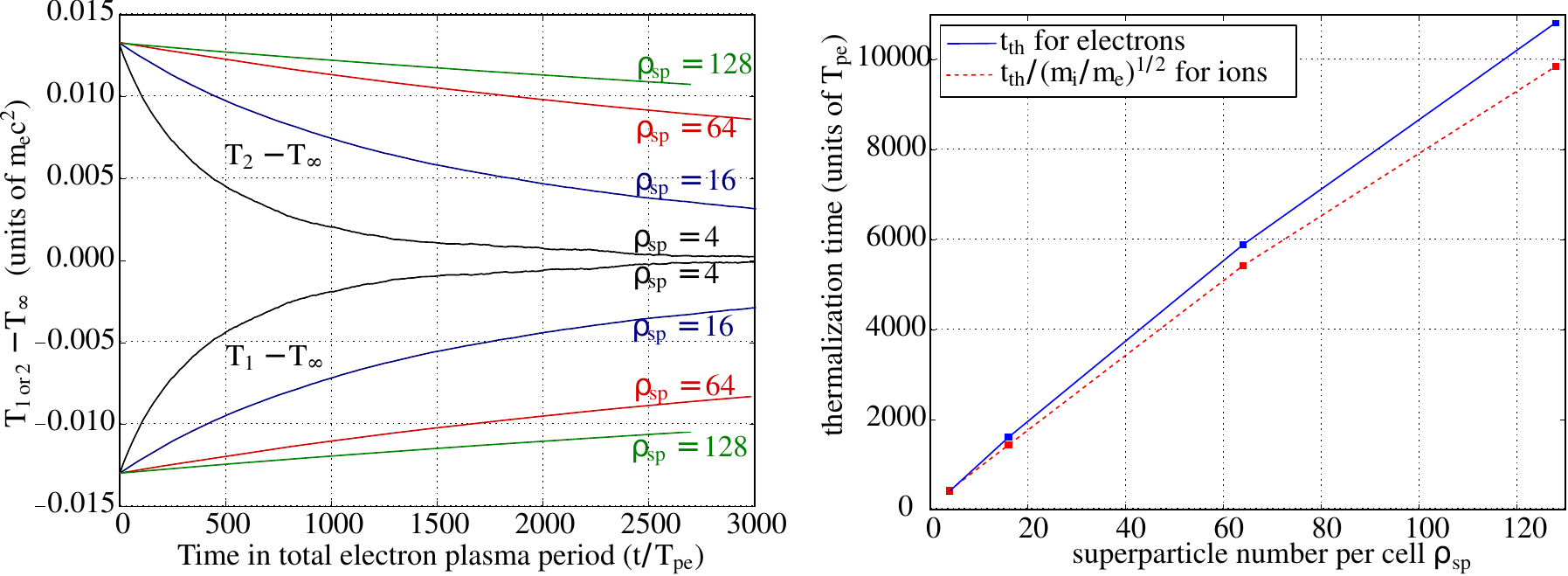}
 \caption{\label{fig:thermalization}%
                 \textbf{Left}: Electron temperatures for the cold (1) and hot (2) plasmas, from four simulations with different 
                 numbers of superparticle per cell $\rho_\mathrm{sp}$ (that include all species).\newline
                 \textbf{Right}: Half-thermalization times for ions and electrons, against number of superparticles per cell. 
 }
\end{figure}

In a PIC plasma, the behavior of plasma quantities depending on $\Lambda$ can be guessed by replacing $\Lambda$
by $\Lambda_p^\mathrm{PIC}$. 
This is the case for the thermalization time of a plasma 
by grazing Coulomb collisions\cite{Spitzer1965} or by electric field fluctuations\cite{Birsdall1985},
which is on the order of $t_\mathrm{th}\sim T_\mathrm{P}\times\Lambda$ (with $T_\mathrm{P}$ the plasma period).
This has two important consequences:
(i) we expect $t^\mathrm{PIC}_\mathrm{th}$ to depend on resolution and coarse-graining, roughly as 
${t^\mathrm{PIC}_\mathrm{th}/T_\mathrm{P}} \propto \Lambda_p = \rho_\mathrm{sp} (\lambda_\mathrm{D} / X_0)^3$,
and (ii) since $\Lambda^\mathrm{PIC}_p = \Lambda/p$ is several orders of magnitude smaller than the real plasma parameter $\Lambda$,
we expect the thermalization due to grazing collisions and fluctuations to be vastly more efficient in PIC codes than in reality.

We present here simulations with initially two thermal ion-electron plasmas (mass ratio $m_\ionperso/m_\lec = 25$) with two different initial temperatures.
The four species interact via collisions and correlations (no sign of plasma kinetic instabilities were found)
and tend to reach the same final temperature $T_\infty$.

Results are shown in Fig.~\ref{fig:thermalization}, for four simulations
with different numbers of superparticles per cell. 
To evaluate the thermalization times we use the exponential variation of the temperature curves\cite{mmelzani}.
We clearly see a slower thermalization as $\rho_\mathrm{sp}$ increases, with a scaling 
$t^\mathrm{PIC}_\mathrm{th} \propto \rho_\mathrm{sp}$ roughly correct for both electrons and ions.

We also emphasize the difference with a real plasma, where $t_\mathrm{th}/T_\mathrm{pe} \sim \Lambda$ 
reaches $10^{10}$ or more, while it is on the order of $\Lambda_p \leq 10^4$ in PIC simulations.

\section{Coarse-Graining and Fluctuation Levels}

\label{Sec:noise_level}

\begin{figure}
 \centering
 \includegraphics[width=\textwidth]{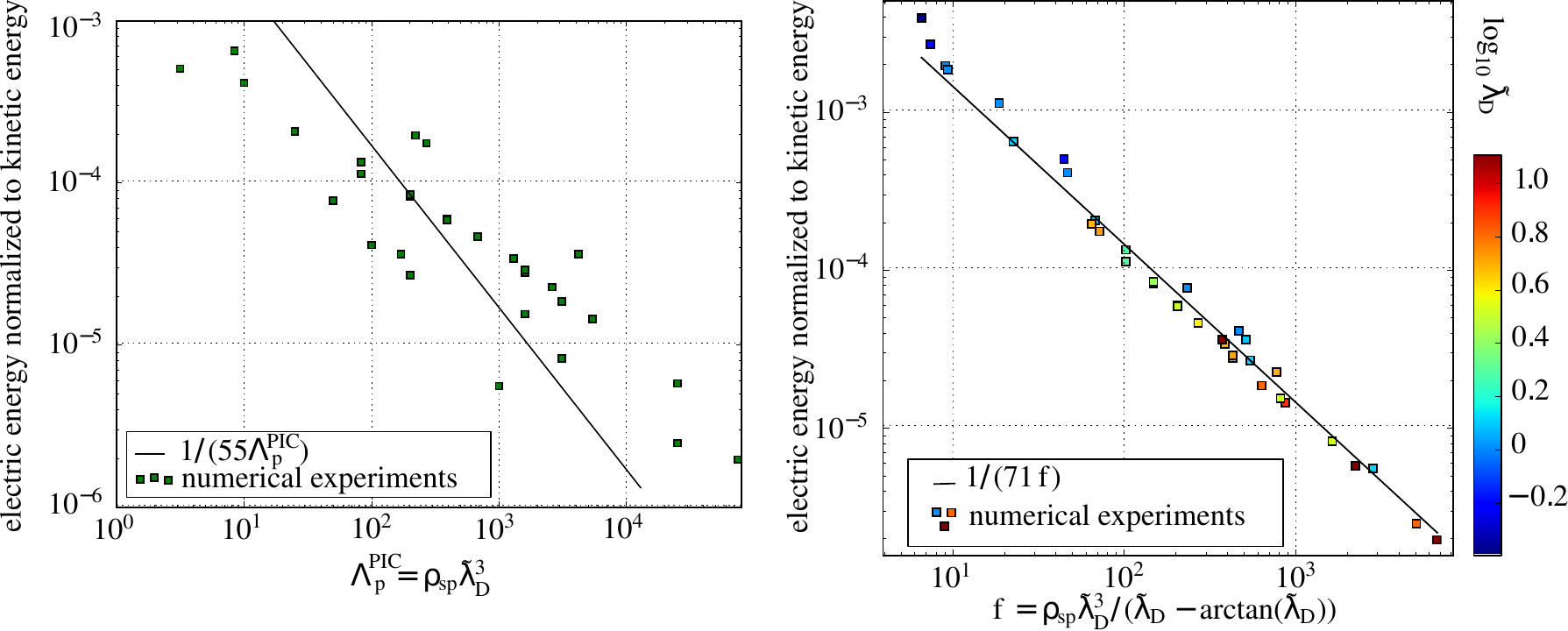}
 \caption{\label{fig:fluctuations}%
  Each point is the measure of the fluctuation level $\varepsilon$ from a PIC simulation.
  Here $\tilde{\lambda}_\mathrm{D} = {\lambda}_\mathrm{D}/X_0$ is the Debye length normalized by the cell size.
 }
\end{figure}

We now study the level of electric field fluctuations in a PIC thermal plasma,
and show that in addition to 
the substitution $\Lambda \rightarrow \Lambda^\mathrm{PIC}_p$, effects due to finite superparticle size must be taken into account.
In a real plasma in thermal equilibrium, it is given by\cite{Callen2006}
\begin{equation}\label{equ:electric_fluctuations_real_plasma}
 \varepsilon = \frac{\langle \epsilon_0 \textbf{E}^2/2\rangle}{3nT/2} \sim \frac{1}{\Lambda},
\end{equation}
where the symbol $\langle\cdot\rangle$ denotes an average over space, $\textbf{E}$ is the electric field, and $n$ and $T$ 
the plasma number density and temperature.

We perform simulations of thermal plasmas at rest and measure the level of energy in the electric field. 
We use thermal velocities from 0.04$c$ to 0.10$c$, $\rho_\mathrm{sp}$ from 2 to 500, and different spatial resolutions.
The results are summarized in Fig.~\ref{fig:fluctuations}, left panel: $\varepsilon$ is not found 
proportional to $1/\Lambda_p^\mathrm{PIC}$.

To explain this, we compute the electric field produced by particles in a thermal plasma\cite{mmelzani}.
The expression for the electric field is a generalization of the Debye electric field for moving particles,
and is integrated in $k$-space from 0 to a maximal wavenumber $k_\mathrm{max}=X_0^{-1}$ given by the superparticle size $X_0$.
We then obtain
\begin{equation}\label{equ:field_fluct_new}
 \frac{\langle \epsilon_0 E_x^2/2\rangle}{3nT/2} 
   = \frac{1}{18\pi^2}\frac{{\lambda}_\mathrm{D}/X_0-\arctan {\lambda}_\mathrm{D}/X_0}{\rho_\mathrm{sp}({\lambda}_\mathrm{D}/X_0)^3},
\end{equation}
in very good agreement with the simulations (Fig.~\ref{fig:fluctuations}, right panel).

We note that for a very high resolution, $\tilde{\lambda}_\mathrm{D} \gg 1$, the field energy 
decreases as $1/(\rho_\mathrm{sp}\tilde{\lambda}_\mathrm{D}^2)$,
which is non-trivial and different from what is expected in a real plasma where 
it decreases as $1/\Lambda = 1/(n{\lambda}_\mathrm{D}^3)$. 

\section{Coarse-Graining and Linear Growth Rates of Instabilities}

\begin{figure}
 \centering
 \includegraphics[width=\columnwidth]{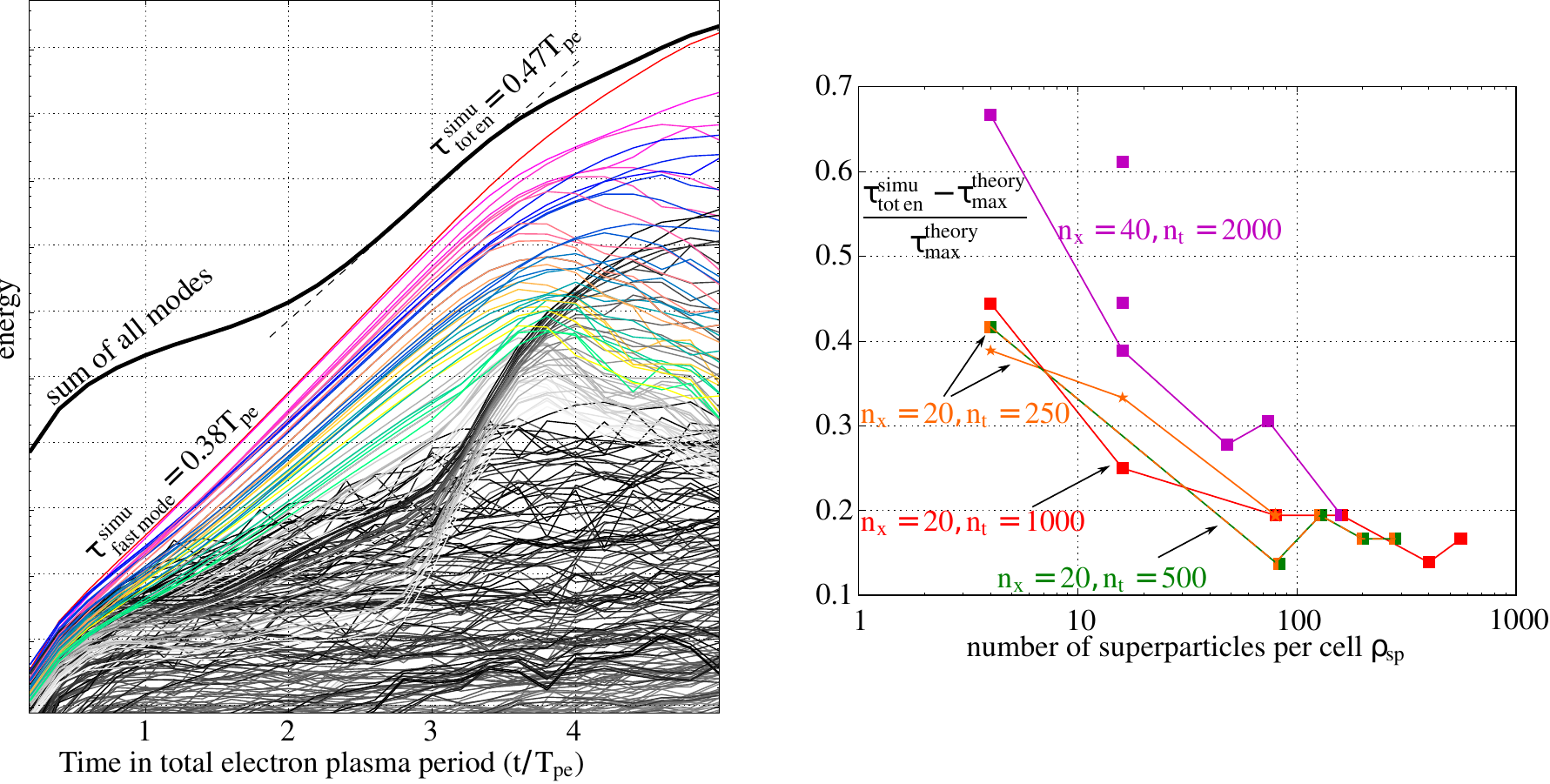}
 \caption{\label{fig:twostream_Fourier_modes}\textbf{Left}: Growth of individual Fourier modes for $b_x$.
                                             \textbf{Right}: Relative difference between the growth rate 
                                             measured on the total energy and the analytical prediction for various simulations.
                                             The timestep is $T_\mathrm{pe}/n_t$ and the grid size is $cT_\mathrm{pe}/(2\pi)$ (with $T_\mathrm{pe}$ the 
                                             electron plasma period).
          }
\end{figure}

This section illustrates the incidence of the high level of noise in the linear phase of instabilities. 
We perform several simulations of cold and unmagnetized counter-streaming beams of pair plasmas, 
with parameters such that the dominant unstable mode is the filamentation instability. 
We measure the growth rate during the linear phase with two methods:  
by a direct measure on the total energy curve, e.g., $\int\!\dif V\, b_x^2 \propto \exp(2t/\tau)$, giving an
effective growth rate that we denote by $\tau^\mathrm{simu}_\mathrm{tot\,en}$,
and by following the time evolution of the Fourier modes of the magnetic field. 

Figure~\ref{fig:twostream_Fourier_modes} (left) is an example of the temporal evolution 
of the modes of $b_x$.
The sum of all modes grows at the same effective growth rate as the total energy 
in $b_x$, $\tau^\mathrm{simu}_\mathrm{tot\,en} = 0.48 T_\mathrm{pe}$.
However, the fastest growing modes grow with $\tau^\mathrm{simu}_\mathrm{fast\,mode} = 0.38T_\mathrm{pe}$, which is 
close to the analytical cold-fluid result $\tau^\mathrm{theory}_\mathrm{max} = 0.36T_\mathrm{pe}$.
We see from Fig.~\ref{fig:twostream_Fourier_modes} (left) that the large difference between the effective growth rate $\tau^\mathrm{simu}_\mathrm{tot\,en}$ 
and the growth rate of the fastest modes $\tau^\mathrm{simu}_\mathrm{fast\,mode}$ 
is due to a significant contribution of all the modes during the whole linear phase. 
The fastest mode thus never dominates the total energy in the linear phase. 

These results hold for all the test simulations that we conducted\cite{mmelzani}: 
the growth rates of the fastest Fourier modes remain at a constant value,
while the effective growth rates measured on 
the total energy present various levels of discrepancies with theory, 
between 14\% and 67\%.
Figure~\ref{fig:twostream_Fourier_modes} (right) shows that there is a systematic decrease in this discrepancy
when the superparticle number per cell $\rho_\mathrm{sp}$ is increased (all other parameters being kept constant).
Since the fluctuation level in the PIC plasma decreases with increasing $\rho_\mathrm{sp}$,
this indicates that the high fluctuation level excites all the modes and prevents the fastest ones from
dominating the energy.

\section{Conclusion}

PIC simulations employ a high level of coarse-graining: each superparticle represents up to $p\sim10^{10}$ real plasma particles.
This would not be an issue for simulating collisionless and correlation-less plasmas because the Vlasov-Maxwell system is invariant under 
coarse-graining. 
However, the physics of collisions and correlations is coarse-graining dependent, and is strongly enhanced by the smallness of the PIC 
plasma parameter $\Lambda_p^\mathrm{PIC} = \Lambda^\mathrm{real}/p \sim \mathrm{a~few}$. 
As a result, noise and fluctuation levels are larger than in the real plasma by a factor $p\sim10^{10}$, and
collision- and fluctuation-induced thermalization is faster by a factor $p\sim10^{10}$.
For example, in real collisionless shocks, 
the mean free path for collisions $l_\mathrm{mean\,free\,path}$ is far larger than the shock thickness $\Delta_\mathrm{shock}$ 
and the thermalization processes are collisionless kinetic instabilities.
Since the mean free path $l^\mathrm{PIC}_\mathrm{mean\,free\,path}\propto\Lambda_p$ 
in a PIC plasma is smaller by a factor of $p\sim 10^{10}$
than in the real plasma, one has to check that
$l^\mathrm{PIC}_\mathrm{mean\,free\,path} \gg \Delta_\mathrm{shock}$ still holds. 
More generally, to truly describe a collisionless and correlation-less plasma with a PIC algorithm, 
one has to be careful that the unphysically enhanced collisional physics remains slower
than the collisionless physics.
This can be achieved with large enough numbers of superparticles per cell, or/and by the use 
of smoother particle shapes that reduce fluctuation levels.

We also showed that the behavior of PIC plasma quantities can be guessed 
by the substitution $\Lambda\rightarrow\Lambda^\mathrm{PIC}_p$, 
but that this recipe is not exact and that the finite size of the superparticles 
is to be taken into account to obtain the precise dependency.


\end{document}